\documentclass{emulateapj}
\usepackage{apjfonts}
\usepackage{bbm}
\usepackage{graphicx}
\usepackage{amssymb,amsmath}
\usepackage{xcolor}
\definecolor{ao}{rgb}{1.0,0.03,0}
\definecolor{ap}{rgb}{0, 0.18,0.39}

\usepackage{color}
\definecolor{light}{rgb}{0.3, 0.3, 0.3}
\def\light#1{{\color{light}#1}}
\usepackage{array}
\hypersetup{breaklinks,colorlinks,linkcolor=ao,citecolor=ap}
\shorttitle{A new ILC method for weak CMB B-mode Signal}

\begin{document}

\title{A Bayesian ILC method for CMB B-mode posterior estimation and reconstruction of primordial gravity wave signal.}

\author{Sarvesh Kumar Yadav\altaffilmark{1},  Rajib Saha\altaffilmark{1}} 

\altaffiltext{1}{Physics Department, Indian Institute of Science 
Education and Research Bhopal,  Bhopal, M.P, 462066, India.}

\begin{abstract}
The Cosmic Microwave Background (CMB) radiation B mode polarization signal contains the unique signature of primordial metric perturbations produced during the inflation. The separation of the weak CMB B-mode signal from strong foreground contamination in observed maps is a complex task, and proposed new generation low noise satellite missions compete with the weak signal level of this gravitational background. In this article, for the first time, we employ a foreground model-independent internal linear combination  (ILC) method to reconstruct the CMB B mode signal using simulated observations over
large angular scales of the sky of  6 frequency bands of future generation CMB mission Probe of Inflation and Cosmic Origins (PICO). We estimate the joint CMB B mode posterior density following the interleaving Gibbs steps of B mode angular power spectrum and cleaned map samples
using the ILC method. We extend and improve the earlier reported Bayesian ILC method to analyze weak CMB B mode reconstruction by introducing noise bias corrections at two stages during the ILC weight estimation. By performing $200$ Monte Carlo simulations of the Bayesian ILC method, we find that our method can reconstruct the CMB signals and the joint posterior density accurately over large angular scales of the sky. We estimate Blackwell-Rao statistics of the marginal density of CMB B mode angular power spectrum and use them to estimate the joint density
of scalar to tensor ratio $r$ and a lensing power spectrum amplitude $A^{\textrm{lens}}$. Using $200$ Monte Carlo simulations of the delensing approach, we find that our method can achieve an unbiased detection of  the primordial gravitational wave signal  $r$ with more than
8$\sigma$ significance for levels of $r \geqslant 0.01$.  
\end{abstract}

\maketitle

\section{Introduction}
\label{sec:intro}

The discovery of the CMB in the second half of the last century led to an era of precision cosmology, which resulted in the demise of many cosmological models and a few's survival. The current observations have put very stringent constraints \citep{2018arXiv180706211P, 2018arXiv180706209P} on the various inflationary models \citep{2014JCAP...03..039M}, within the very successful inflationary paradigm for the origin of the primordial perturbations. The final Planck 2018 release \citep{2018arXiv180706211P}, ruled out the perfect scale invariance for the spectral index of scalar perturbations at 8.4 $\sigma$ and the running and the running of the running \citep{2003PhRvD..68f3501C}, of the spectral index have been negated with 95 \% CL, consistent with the simplest slow-roll dynamics for the inflaton, and the spatial curvature $\Omega_{K}$ is $ -0.011_{-0.012}^{+0.013}$ at 95 \% CL. BICEP2/Keck Array \citep{2016ApJ...833..228B} together with Planck 2018 strongly disfavors monomial models with $V(\phi) \propto \phi^{p}$, $p>1$, natural inflation, and low scale SUSY models. The observations have also established that the primordial perturbations are adiabatic to a very high degree \citep{2018arXiv180706211P}, and the primordial power spectrum does not deviate from a pure power-law \citep{2018arXiv180706211P}. Further the Planck likelihood together with the B-mode polarization likelihood of the BICEP2-Keck Array puts a stringent 95 \% CL upper limit of $r_{0.002} < 0.056$ corresponding to the energy scale of inflation of $V^{1/4} < 1.6 \times 10^{16}$ GeV with 95 \% CL bound.

To further constraint the cosmological models and probe the energy scale of the inflation and the existence of primordial gravitational waves, new generation of CMB space missions such as PICO \citep{2019BAAS...51g.194H} COrE \citep{2018JCAP...04..014D}, LiteBIRD \citep{2014JLTP..176..733M} and PIXIE \citep{2011JCAP...07..025K}  have been proposed to detect the primordial CMB B mode on large angular scales at a level of $r < 10^{-3}$. The CMB B-mode signal given the current bound will have typical RMS fluctuations $<0.1$ $ \mu K$, which are extremely weak, with the strong polarized Galactic foregrounds and instrumental systematic making their detection and reconstruction extremely difficult. To further add to difficulties, gravitational lensing introduces spurious cosmic variance from the lens-induced B modes or the lensing B modes. The lensing B-modes are due to the conversion of CMB E-mode to B-mode due to weak gravitational lenses along the line of sight \citep{2004PhRvD..69d3005S}. Not only this biases the amplitude $r$ but also the cosmic variance of the primordial CMB B-mode power spectrum.


Over the recent years, various studies dealing with the foregrounds minimization in the context of CMB B-mode sky has been undertaken \citep{2004Bachhigalupi,2009A&A...503..691B,2009AIPC.1141..222D,2011MNRAS.414..615B,2011ApJ...737...78K,2012MNRAS.424.1914A,2016JCAP...03..052E,2016MNRAS.458.2032R,2018JCAP...04..023R,2017MNRAS.468.4408H}. Recently, new methods were proposed to investigate the joint posterior density of CMB signal and corresponding theoretical angular power spectrum on large scales, for CMB temperature \citep{2018ApJ...867...74S,2018arXiv181008872S} and CMB E-mode \citep{2020arXiv200313570P,purkayastha2020foreground} polarization. 

 In this article, we extend \citep{2018arXiv181008872S,purkayastha2020foreground} and develop a new non-parametric method by also taking care of detector noise to reconstruct clean CMB B-mode signal and corresponding theoretical angular power spectrum.
  We perform a joint analysis of CMB B-mode signal and its angular power spectrum posterior density without considering any foreground model. Unlike the polarized CMB, we do not have an accurate enough model for the polarized galactic foreground and the exact number of independent polarized foregrounds is not known \citep{2016MNRAS.458.2032R}. Our non-parametric method avoid effects \citep{2012MNRAS.424.1914A} due to inaccurate polarized Galactic foreground models. Our method also provides the best fit estimates of both, CMB B-mode map and it's theoretical angular power spectrum along with their confidence interval regions. We apply our Bayesian ILC method to recover the weak CMB B-mode signal from the simulated foreground and noise-contaminated 6 PICO frequency channels. We look into the performance of our Bayesian ILC method following the Gibb's procedure to reconstruct the primordial CMB B-mode signal and its theoretical angular power spectrum. We also perform correction for lensing bias in CMB B-mode power spectrum without removing the lensing cosmic variance contribution to B-modes. We use the samples of the theoretical CMB B-mode power spectrum generated at each Gibb's step of our method and simultaneously fit the amplitude of the primordial B-mode power spectrum parameter and the amplitude of lensing B-mode power spectrum in a Bayesian framework \citep{2018MNRAS.474.3889R}. The method removes the lensing bias on the posterior distribution of $r$ and enables us to detect $r$ with more than $8\sigma$ significance for CMB B-mode satellite mission like PICO for $r\ge 0.01$.

We organize our paper as follows. In Section \ref{formalism}, we illustrate the basic formalism of this work by describing our algorithm used to get the clean sky, angular power spectrum along with their Gibb's samples. In Section \ref{simulations}, we describe the procedure to get the foreground and noise-contaminated B-mode maps at 6 PICO frequencies. In Section, \ref{method} we discuss the method adopted to get the samples of reconstructed CMB B-mode map, theoretical angular power spectrum along with the delensing procedure. In Section \ref{result}, we first present and discuss results obtained for the cleaned map, then for the angular power spectrum, and then we discuss and present results for the delensing technique using Blackwell-Rao approximation \citep{2005PhRvD..71j3002C} to obtain the unbiased posterior distribution of $r$. Finally, in Section \ref{Diss}, we discuss and conclude.

\section{Formalism} 
 \label{formalism}
 
 This section discusses the formalism used to estimate the joint posterior density of the CMB signal and its theoretical angular power spectrum given the observed data. We adopt and improve formalism used in this work for component separation as in \citep{2018arXiv181008872S,purkayastha2020foreground} so that it applies to the weak CMB B-mode signal. The method not only gives us the best-fit CMB B-mode map and it is best-fit power spectrum, but it also provides MCMC Gibbs samples for sky power spectrum $\hat{\sigma}_{\ell}^{B}$ and theoretical power spectrum $\textit{C}_{\ell}^{B}$, which we utilize to delens the angular power spectrum and to estimate of tensor-to-scalar ratio $r$.
 
 \subsection{Data Model}
 \label{DM}
 Given observations of CMB B-mode signal \textbf{S} at $n$ different frequencies in thermodynamic temperature units, we can write for an observed $\textit{i}^{ th}$ frequency map  $\textbf{X}_{i}$,
 
 \begin{equation}
 \textbf{X}_{i} = \textbf{S} + \textbf{F}_{i} + \textbf{N}_{i}
 \label{eq1}
 \end{equation}
 where $\textbf{F}_{i}$ is the net foreground contribution from all the foreground components at the $\textit{i}^{ th}$ frequency channel and $\textbf{N}_{i}$ is the corresponding detector noise. Each of the above bold-faced quantity is a column vector of size $N_{pix}$ representing a HEALPix\footnote{Hierarchical Equal Area Isolatitude Pixellization of sphere, e.g., see	\cite{2005ApJ...622..759G}} map where $N_{pix}=12{N_{side}}^{2}$, $N_{side}$ being the  pixel resolution parameter, having common beam and pixel resolution. Let \textbf{D} denote the observed data set i.e. $\textbf{D} = \{\textbf{X}_{1},\textbf{X}_{2},...,\textbf{X}_{n}\}$.
 
 \subsection{CMB Posterior Estimation}
 \label{GSCMBP}
 Given the observed data, \textbf{D}, $P(\textbf{S},\textit{C}_{\ell}^B|\textbf{D})$ represents the joint density of CMB B-mode map, \textbf{S}, and the theoretical  CMB  B mode angular power spectrum, $\textit{C}_{\ell}^B$. As it is difficult to obtain the $P(\textbf{S},\textit{C}_{\ell}^B|\textbf{D})$ analytically  we evaluate it by drawing samples from it. If we can sample from the conditional distributions $P(\textbf{S}|\textit{C}_{\ell}^B,\textbf{D})$ and $P(\textit{C}_{\ell}^B|\textbf{S},\textbf{D})$ then utilizing Gibbs sampling approach~\cite{GR1992}, which says that samples $(\textbf{S}^{i}, \textit{C}_{\ell}^{B~i})$ can be drawn from the joint distribution $P(\textbf{S},\textit{C}_{\ell}^B|\textbf{D})$ by iterating the following symbolic sampling equations:
 \begin{equation}
 \textbf{S}^{i+1} \leftarrow P(\textbf{S}|\textit{C}_{\ell}^{B~i},\textbf{D})
 \label{eq2}
 \end{equation}
 
 \begin{equation}
 \textit{C}^{B~i+1}_{\ell} \leftarrow P(\textit{C}_{\ell}^B|\textbf{S}^{i+1})
 \label{eq3}
 \end{equation}
  The symbol ``$\leftarrow$'' implies that a sample of corresponding variables is drawn from the distribution on the right-hand side. Once the initial burn-in period is over, the samples will converge to being drawn from the required joint distribution. 
 
 \subsubsection{Sampling CMB Signal}
 \label{STCMBS}
We use foreground model-independent method to draw samples of \textbf{S} given the CMB B-mode theory $\textit{C}_{\ell}^B$ and \textbf{D}. We modify the global ILC method described in \citep{Sudevan_2018,purkayastha2020foreground} to improve separating the weak CMB B-mode signal given the detector noise model. Let us assume that the mean corresponding to each frequency map $\textbf{X}_{i}$, as discussed in (\ref{DM}), has already been subtracted. The cleaned CMB B mode map $\textbf{S}$ can be obtained by linear combination of $n$ input maps $\textbf{X}_{i}$, with weight factor $\textit{w}_{i}$, i.e.,
 \begin{equation}
 \textbf{S} = \sum_{i=1}^{n} \textit{w}_{i}\textbf{X}_{i}.
 \label{eq4}
 \end{equation}
Since the spectral distribution of CMB photons is a blackbody to an excellent approximation, the CMB anisotropy signal \textbf{S} (in thermodynamic temperature units) is independent of the frequency channel. In order to avoid multiplicative bias in amplitudes of CMB anisotropies the sum of weights is constraint to unity i.e., $\sum_{i=1}^{n} w_{i} = 1$. 
As discussed in \citep{Sudevan_2018} instead of minimizing the clean map variance $\textbf{S}^{T}\textbf{S}$ we minimize   
 \begin{equation}
 \sigma^{2} = \textbf{S}^{T}\textbf{C}^{\dagger}\textbf{S}
 \label{eq5}
 \end{equation}
 where $\textbf{C}$ represents the CMB B-mode theoretical covariance matrix and $^{\dagger}$ denotes the Moore-Penrose generalized inverse \cite{penrose_1955}. Using Equation~(\ref{eq4}) in Equation~(\ref{eq5}) we write,
 \begin{equation}
 \sigma^{2} = \textbf{W}\textbf{A}\textbf{W}^{T}
 \label{eq6}
 \end{equation}
 where $\textbf{W} = (w_{1},w_{2},w_{3},...,w_{n})$ is a $1\times n$ weight row vector and \textbf{A} is an $n \times n$ matrix with it element $A_{ij}$ given by 
 \begin{equation}
 A_{ij} = \textbf{X}_{i}^{T}\textbf{C}^{\dagger}\textbf{X}_{j}
 \label{eq7}
 \end{equation} 
 
 The weights that minimize the variance given by Equation~(\ref{eq6}) subject to the above constraint is obtained following  Lagrange's multiplier approach \citep{2008PhRvD..78b3003S,1996MNRAS.281.1297T,2006ApJ...645L..89S,2003PhRvD..68l3523T} and is given by 
 \begin{equation}
 \textbf{W} = \frac{\textbf{e} \textbf{A}^{\dagger}}{\textbf{e} \textbf{A}^{\dagger}\textbf{e}^{T}}
 \label{eq8}
 \end{equation} 
 where $\textbf{e} =(1, 1 ,..., 1)$ is the $1 \times n$ CMB shape vector in thermodynamic temperature units and $\textbf{A}^{\dagger}$ is the Moore-Penrose generalized inverse of the matrix \textbf{A}. Computing a dense matrix $[\textbf{C}]_{N_{pix} \times N_{pix}}$ at every Gibbs iteration is computationally costly, hence we switch to the harmonic space where Equation (\ref{eq7}) is simpler to compute,
 \begin{equation}
 A_{ij} = \sum_{\ell = 2}^{l_{max}}(2\ell +1)\frac{\sigma_{\ell}^{ij}}{C_{\ell}^{B}},
 \label{eq9}
 \end{equation} 
 where $\ell_{max}$ denotes the maximum multipole used in he analysis, $\sigma_{\ell}^{ij}$ denotes the angular cross power spectrum between $\textbf{X}_{i}$ and  $\textbf{X}_{j}$ channel maps and $C_{\ell}^{B}$ represents the beam and pixel smoothed CMB BB theoretical power spectrum i.e., 
 \begin{equation}
 C_{\ell}^{B}=C_{\ell}^{B'}B_{\ell}^{2}P_{\ell}^{2}
 \label{eq10}
 \end{equation}
 where $C_{\ell}^{B'}$ does not have any smoothing effect, and $B_{\ell}$ and $P_{\ell}$ are respectively the polarization beam and polarization pixel window functions. The internal linear combination method for component separation performs well only in a low noise environment, and since the CMB B-mode signal is even weaker than CMB E mode by order of magnitude, to minimize the residual noise bias in the output CMB-B mode map and power spectrum  we subtract the noise auto-power initially from the input frequency cross power spectrum,
 
 \begin{equation}
 A_{ij} = \sum_{\ell = 2}^{l_{max}}(2\ell +1)\frac{1}{C_{\ell}^{B}}(\sigma_{\ell}^{ij}-\delta_{ij}\sigma_{\ell}^{N,i}).
 \label{eq11}
 \end{equation}
 where  $\sigma_{\ell}^{N,i}$ is noise auto power corresponding to detector at $i^{th}$ frequency. 
 We use matrix $\textbf{A}$, the component for which are given by Equation (\ref{eq11}), in Equation (\ref{eq8}) to  obtain the row vector \textbf{W}. We use the weights obtained, to sample the foreground minimized CMB B-mode signal \textbf{S}, by linearly combining the input channel maps $\textbf{X}_{i}$ at every Gibb's step following the Equation (\ref{eq4}).   
 
 \subsubsection{Sampling $C_{\ell}^{B}$}
 The signal sample \textbf{S} can be represented mathematically in terms of spherical harmonics,
 \begin{equation}
 \textbf{S}(\theta,\phi) = \sum_{\ell=2}^{\infty}\sum_{m=-\ell}^{\ell}s_{\ell m}\textbf{Y}_{\ell m}(\theta,\phi),
 \label{eq12}
 \end{equation}
 then the realization-specific power spectrum  is given by
 \begin{equation}
 \hat{\sigma}_{\ell}^{B'} \equiv \frac{1}{2\ell + 1}\sum_{m=-\ell}^{\ell}|s_{\ell m}|^{2}.
 \label{eq13}
 \end{equation}
 In order to minimize the noise bias in the sampled theory $C_{\ell}^{B}$ we further subtract the weighted noise power from $\hat{\sigma}_{\ell}^{B'}$ to obtain,
 \begin{equation}
 \hat{\sigma}_{\ell}^{B} = \hat{\sigma}_{\ell}^{B'} - \sum_{i=1}^{n}w_{i}^{2}\sigma_{\ell}^{N,i}
 \label{eq14}
 \end{equation}
 Since the power spectrum only depends on the signal \textbf{S} through $\hat{\sigma}_{\ell}^{B}$, and not its phases, 
 therefore to draw samples of $C_{\ell}^{B}$ given \textbf{S}, we sample from $P(\textit{C}_{\ell}^B|\hat{\sigma}_{\ell}^{B})$.
 The conditional density $P(\textit{C}_{\ell}^B|\hat{\sigma}_{\ell}^{B})$ can be written \cite{2018arXiv181008872S} as, 
 \begin{equation}
 P(C_{\ell}^{B}|\hat{\sigma}_{\ell}^{B}) = \left(\frac{1}{C_{\ell}^{B}}\right)^{(2\ell + 1)/2} \exp\left[-\frac{\hat{\sigma}_{\ell}^{B}(2\ell + 1)}{2C_{\ell}^{B}}\right],
 \label{eq15}
 \end{equation}
 where the variable $x = \hat{C}_{\ell}^{B}(2\ell +1)/C_{\ell}^{B}$ is a $\chi^{2}$ distributed random variable having $2\ell-1$ degrees of freedom. In order to draw samples of $C_{\ell}^{B}$ using Equation (\ref{eq14}) we need to draw first x from the $\chi^{2}$ distribution of $2\ell-1$ degrees of freedom. For this we draw $2\ell-1$ independent normal variables and then sum their squares. Therefore given \textbf{S} we  have estimates of $\hat{\sigma}_{\ell}^{B}$, we then obtain $C_{\ell}^{B}$ using $C_{\ell}^{B} = \hat{\sigma}_{\ell}^{B}(2\ell+1)/x$. 
 
  \begin{figure*}
 	\hspace{1.cm}
 	\includegraphics[scale=.9]{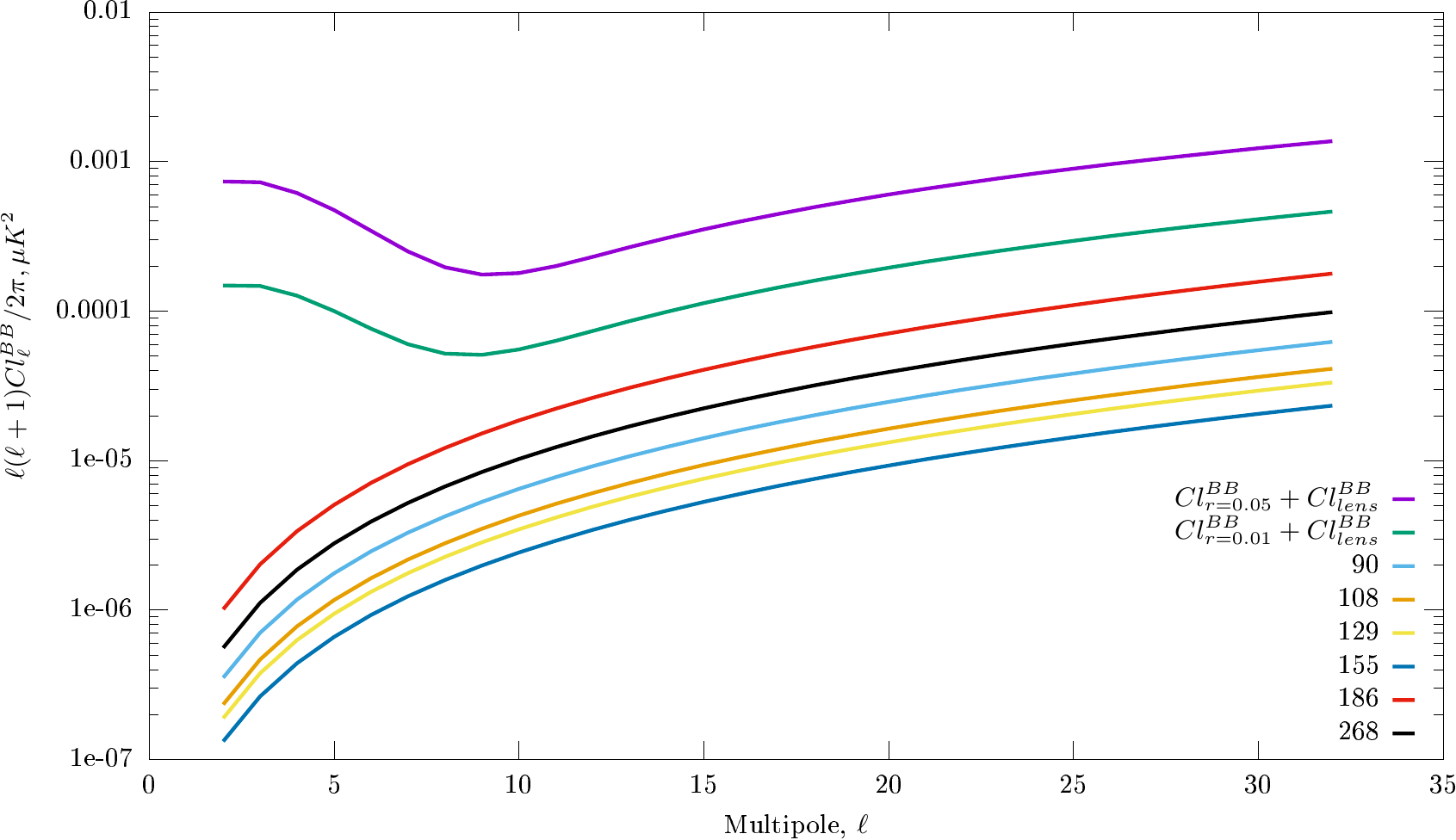}
 	\caption{The above plot shows the noise power spectrum corresponding to each of the six PICO frequencies used in this work, along with the lensed B-mode power spectrum for $r=0.01$ and $r=0.05$. From the figure, we find that the CMB B-mode theoretical angular power spectrum is well above the noise power for the 6 PICO frequency bands used in this work.}
 	\label{fig:noise-crop}
 \end{figure*}

 \section{FREQUENCY MAPS}
 \label{simulations} 
 
 In this work, we simulate the foreground and noise-contaminated CMB B modes at 6 CMB dominating the least noisy frequency bands of proposed satellite mission PICO, in the frequency range 90 GHz to 268 GHz. We list the frequency channel maps along with their instrumental specifications in Table (\ref{table:picofreq}).

 \begin{table}
 	\centering
 	\begin{tabular}{lll}
 		\hline
 		Frequency & Beam FWHM & Q and U noise RMS            \\
 		(GHz)     &   (arcmin)     & ($\mu$$K_{CMB}$ $arcmin$)           \\  
 		\hline
 		\vspace*{-.7em}
  		~~~   & ~~~~ & ~~~~~~~~~ \\		
 		~~~~~90   & ~~~~9.5 & ~~~~~~~~~2.09 \\
 		~~~108   & ~~~~7.9 & ~~~~~~~~~1.70  \\
 		~~~129  & ~~~~7.4 & ~~~~~~~~~1.53 \\
 		~~~155  &  ~~~~6.2 & ~~~~~~~~~1.28 \\
 		~~~186 &  ~~~~4.3 & ~~~~~~~~~3.54 \\
 		~~~268  & ~~~~3.2 &  ~~~~~~~~~2.63  \\ 
 		\hline
 	\end{tabular}
 	\caption{ PICO frequency maps used in this work along with their instrumental specifications.}
 	\label{table:picofreq}
 \end{table}
 
 \subsection{CMB B-mode Signal}
 
 We simulate lensed CMB $Q$ and $U$ Stoke parameter maps from the lensed CMB B-mode angular power spectra generated by the Boltzmann solver \light{CAMB} \citep{2000ApJ...538..473L}. Since we perform our analysis on large scales ($\ell\le32 $), the non-Gaussianity of lensing B-mode fluctuations \citep{2004PhRvD..70d3002S} can be neglected with respect to primordial Gaussian B-mode fluctuations \citep{2012JCAP...06..014S}. Therefore the likelihood Equation (\ref{eq6}) is relevant for our current work \citep{2018MNRAS.474.3889R}. We performed our analysis on 0.01 and 0.05 tensor-to-scalar values assuming $\lambda CDM + r$ cosmology with optical depth to reionization $\tau=0.055$ \citep{2016A&A...596A.107P}, $A^{\textrm{lens}}=1$ and other cosmological parameters set to the Planck 2015 best-fit values \citep{2016A&A...594A...1P}.
 
 \subsection{Foreground B-mode Signal}
 We generate foreground maps at all the six PICO frequencies used in this work corresponding to synchrotron and thermal dust; two major CMB polarized foreground contributors. To generate them, we follow the procedure similar to \citep{2018MNRAS.474.3889R}. We generate $Q$ and $U$ maps at each frequency and use them to obtain corresponding B-mode maps for both the foregrounds.
 
 To generate the polarized Galactic synchrotron Stokes maps, we extrapolate the \textit{Wilkinson  Microwave Anisotropy Probe} (WMAP) 23 GHz \citep{2007ApJS..170..335P} \citep{2013ApJS..208...20B} stokes maps $Q_{23}$ and $U_{23}$ to the six PICO frequencies through a power-law frequency dependence:
 \begin{equation}
 Q_{\nu}^{sync}(p) = Q_{23}(p)\left(\frac{\nu}{23~GHz}\right)^{\beta_{s}}
 \end{equation}
 
 \begin{equation}
 U_{\nu}^{sync}(p) = U_{23}(p)\left(\frac{\nu}{23~GHz}\right)^{\beta_{s}}
 \end{equation}
 We use a constant spectral index $\beta_{s}=-3$ which is close to the typical mean values measured at CMB frequencies \citep{2016A&A...594A..10P,2013ApJS..208...20B,2009ApJ...705.1607D,2008A&A...490.1093M,2007ApJ...665..355K,1996MNRAS.278..925D} and $p$ is the pixel index.
 
 To simulate the Galactic polarized thermal dust Stokes maps, we extrapolate the generalized needlet ILC (GNILC) Planck 353 GHz thermal dust optical depth map \citep{2016A&A...596A.109P} to the relevant PICO frequencies:
 \begin{equation}
 Q_{\nu}^{dust}(p) = f_{d}g_{d}(p)I_{\nu}^{GNILC}(p)\cos(2\gamma_{d}(p))
 \end{equation}
 \begin{equation}
 U_{\nu}^{dust}(p) = f_{d}g_{d}(p)I_{\nu}^{GNILC}(p)\sin(2\gamma_{d}(p))
 \end{equation}
 where $f_{d}$ is the pixel independent intrinsic dust polarization fraction which depends on the level of depolarization along the line of sight, following \citep{2013A&A...553A..96D,2018MNRAS.474.3889R}  we take it to be 0.15, $g_{d}$ is the pixel dependent geometric depolarization factor which we compute using the 3D Galactic magnetic field and 3D distribution along the line of sight. To compute polarization angle $\gamma_{d}$ \citep{2013A&A...553A..96D} at each pixel, we use WMAP 23 GHz map after smoothing with Gaussian beam of $3^{\circ}$,
 \begin{equation}
 \gamma_{d}(p) = \frac{1}{2}\tan^{-1}\left(\frac{-U_{23}(p)}{~~Q_{23}(p)}\right).
 \end{equation}
 We compute the depolarization factor $g_{d}$ using WMAP 23 GHz, and the residual monopole subtracted 408 MHz Haslam synchrotron template ($I_{(0.408)}$), extrapolated to 23 GHz assuming a constant spectral index of -3.0. To compute it, we smooth the extrapolated map to gauss beam of $3^{\circ}$ at $N_{side}=512$ and use,
 \begin{equation}
 g_{d}(p) = \frac{\sqrt{Q_{23}^{2}(p) + U_{23}^{2}(p)}}{f_{s}I_{(0.408)}(p)(23.0/0.408)^{-3.0}},
 \end{equation} 
where for the spectral index used in above equation, the synchrotron polarization fraction $f_{s}=0.75$. The  $I_{\nu}^{GNILC}$ is the \light{GNILC} dust intensity map free from the cosmic infrared background at the frequency $\nu$ and is given by the modified blackbody spectrum: 
 \begin{equation}
 I_{\nu}^{GNILC}(p) = \tau_{353}^{GNILC}(p) \left(\frac{\nu}{353 GHz}\right) ^{\beta_{d}}B_{\nu}(T_{d})
 \end{equation}
 where $\tau_{353}^{GNILC}$ is the Planck \light{GNLIC} dust optical depth at 353 GHz, the dust emissivity $\beta_{d}=1.6$ and $T_{d}=19.4K$ is the dust temperature. $B_{\nu}(T_{d})$ is the Planck function at thermal dust temperature $T_{d}$ given by:
 \begin{equation}
 B_{\nu}(T_{d}) = \frac{2h^{3}}{c^{2}}\frac{1}{\exp\left(\frac{h\nu}{_{B}T_{d}}\right)-1}
 \end{equation}
 
  We use the above-obtained synchrotron and thermal dust Stokes $Q$ and $U$ maps to get B-mode synchrotron and thermal dust foreground maps at each of the PICO frequencies used in this article at $N_{side}=16$. We smooth the obtained maps by polarized Gaussian beam of FWHM $9^{\circ}$. 
 \subsection{Detector Noise Simulations}
 We simulate Gaussian, isotropic, and pixel-pixel uncorrelated random realizations of detector $Q$ and $U$ noise maps for the six PICO \citep{2018SPIE10698E..46Y} frequency bands used in this work. We present detector specifications for each of the bands in Table (\ref{table:picofreq}). We further assume that $Q$ and $U$ noise maps are pixel uncorrelated i.e.
 \begin{equation}
 \left\langle Q_{i}(p)U_{i}(p')\right\rangle = 0.
 \end{equation} 
 We further assume that the pixel noise variances $\sigma_{Q_{i}}^{2}$ for $Q$ and $\sigma_{U_{i}}^{2}$ for $U$ maps at a frequency $\nu_{i}$ are identical and given by 
 \begin{equation}
 \sigma_{Q_{i}}^{2}=\sigma_{U_{i}}^{2}= \left(c\Delta Q_{i}^{2}\right)^{2}/(\Delta\Omega)
 \end{equation} 

  \hspace{-0.45cm} where $\Delta Q_{i}$ is the noise RMS in arcminute for $Q_{i}$ map, $c$ is the conversion factor from arcminute to radian and $\Delta\Omega$ is the solid angle subtended by single-pixel at $N_{side=16}$. We bring both $Q$ and $U$ noise maps to the same beam resolution at $N_{side}=16$  by multiplying the ratio of a polarized Gaussian beam of FWHM $9^{\circ}$ and the polarized beam is given in Table (\ref{table:picofreq}) for corresponding frequency channel. We finally convert the noise Stokes maps obtained to full sky B-mode noise map at each of the frequencies. 
 In Figure (\ref{fig:noise-crop}) we show the PICO detector model noise power corresponding to the six channels along with the lensed CMB B-mode theoretical power for $r=0.01$ and $r=0.05$. We can see that the noise power for all the used frequency is well below the B-mode signal power spectrum at all multipoles used in this work.   
 Finally, we obtain the simulated PICO input noisy foreground contaminated CMB B-mode maps by combining all the three components for each of the six frequencies using equation (\ref{eq1}).
 
 \begin{figure*}
 	\hspace{1.5cm}
 	\includegraphics[scale=.1]{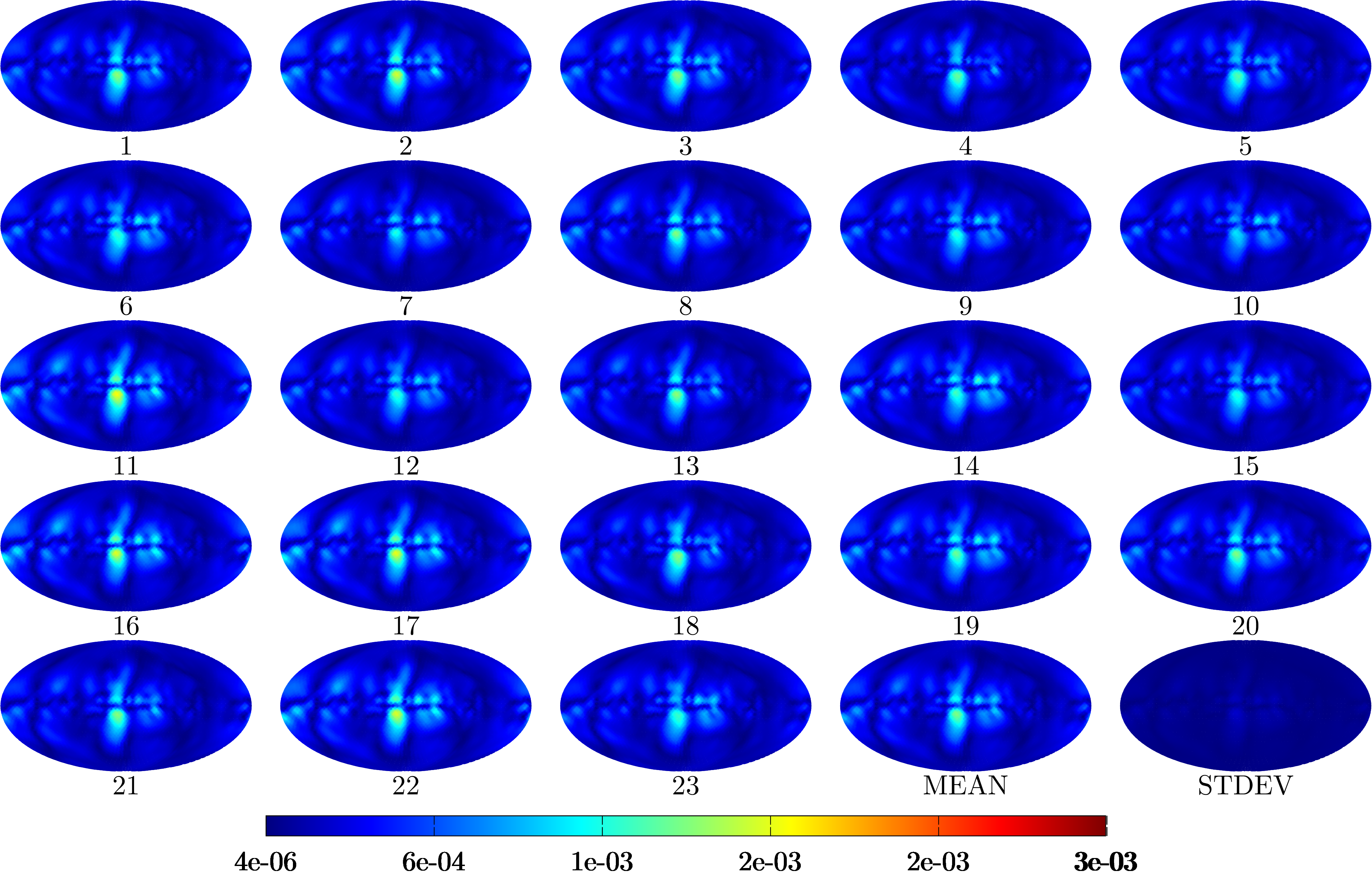}
 	\caption{Above figure shows standard deviation maps for seeds 1 to 23 obtained using CMB signal reconstruction method discussed in this article. The last map shown at the right bottom corner of the figure, labeled as STDEV, shows standard deviation map obtained using the 200 standard deviation maps from the 200 simulations. The second last map shown in the last row of the figure, labeled as MEAN, shows the mean of all the 200 standard deviation maps obtained from the 200 simulations. In all of the above maps, the standard deviation's maximum value is well within the $10^{3} \mu K $, which indicates accurate CMB signal reconstruction. The unit is in $\mu K$ thermodynamic.}
 	\label{fig:sdmulti_r0p05-crop}
 \end{figure*}

\begin{figure*}
	\hspace{1.5cm}
	\includegraphics[scale=.1]{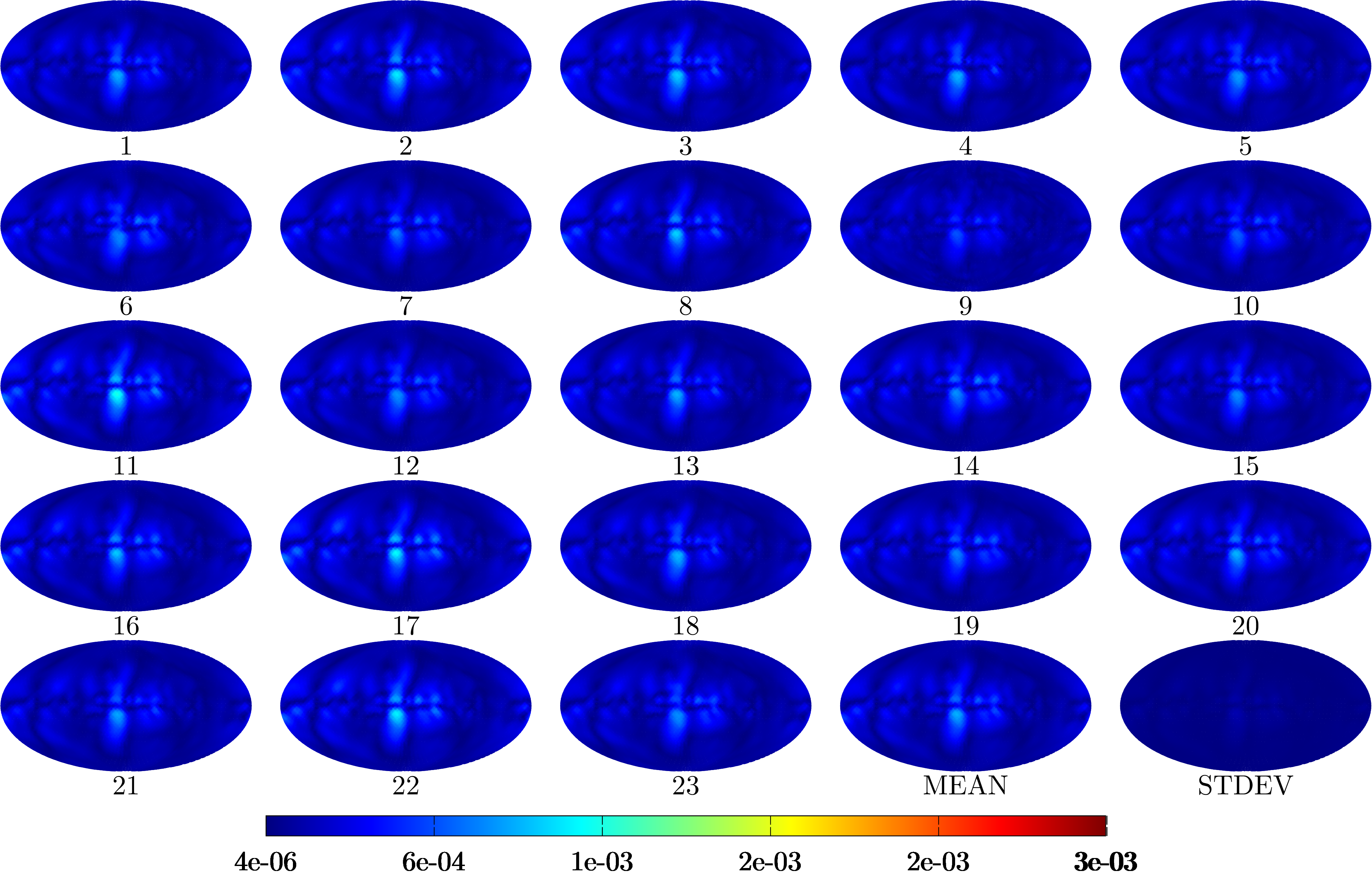}
	\caption{The figure shows 25 standard deviation maps for $r=0.01$. The figure shows standard deviation maps for seeds 1 to 23 obtained using CMB signal reconstruction method discussed in this article. The last map shown at the right bottom corner of the figure, labeled as STDEV, shows standard deviation map obtained using the 200 standard deviation maps from the 200 simulations. The second last map shown in the last row of the figure, labeled as MEAN, shows the mean of all the 200 standard deviation maps obtained from the 200 simulations. All other maps in the above figure are standard deviation map for randomly chosen seeds out of the 200 simulations. In all these maps, the standard deviation's maximum value is well within the $10^{-3} \mu K $, which indicates accurate reconstructions of the CMB signal. The unit is in $\mu K$ thermodynamic.}
	\label{fig:sdmulti_r0p01-crop}
\end{figure*}



\section{Methodology}
 \label{method} 
 
 We implement our model-independent method on the simulated foreground and noise-contaminated B-mode maps obtained above after removing the monopole and dipole components from each of them. We smooth theoretical  $C_{\ell}^{B}$ obtained from CAMB using $\tau=0.055$, $A^{\textrm{lens}}=1$ and Planck 2015 best-fit values by Gauss beam of $9^{\circ}$ and polarization pixel window function corresponding to $N_{side}=16$ as in equation (\ref{eq10}). In order to obtain sampled CMB B-mode sky \textbf{S} given the observed data set \textbf{D} and sampled theory $C_{\ell}^{B~i}$ following the symbolic sampling equation (\ref{eq2}) we first obtain \textbf{A} matrix following equation (\ref{eq11}) and use it to obtain weights using equation (\ref{eq8}). We use weights obtain in the last step to combine the input foreground linearly, and noise-contaminated CMB B-mode maps to obtain the cleaned CMB B-mode map. We obtain sky power $\hat{\sigma}_{\ell}^{B'}$ from the above cleaned CMB B-mode map. We use equation (\ref{eq14}) to obtain weighted noise power subtracted sky power $\hat{\sigma}_{\ell}^{B}$ which is used to sample the theory $C_{\ell}^{B}$. We use ten independent chains; each chain consists of 10000 Gibbs steps. We discard the initial 50 samples for the burn-in period in each chain. In total we obtain 99500 samples of $C_{\ell}^{B}$ and  \textbf{S}. We perform this analysis on cases with $0.01$ and $0.05$ tensor-to-scalar ratios. 

 Using the samples obtained after applying our method, we forecast the proposed CMB space mission PICO's ability to constrain $r$ in the presence of realistic lensing and foreground contributions. We simulate 200 different noise and foreground contaminated Gaussian random CMB B-mode realizations as described in section (\ref{simulations})  and apply Gibb's ILC method to obtain 99500 samples of $C_{\ell}^{B}$ and  \textbf{S} for each of them.
We use a set of samples $\{C_{\ell}^{B~i}\}$ to compute the posterior distribution of the tensor-to-scalar ratio, $r$, and the amplitude of lensing, $A^{\textrm{lens}}$, using the Blackwell-Rao estimator \citep{2005PhRvD..71j3002C} for each of the 200 cases. We use sampled $C_{\ell}^{B~i}$ to obtain the best-fit value of the power spectrum for all the 200 simulations and use them to study bias in the recovered power spectrum. We also obtain a mean map and study reconstruction error in recovered CMB B-mode maps using our method.

\section{Results} 
\label{result}

  This section presents results obtained after applying our method on the simulated foreground and noise-contaminated CMB B-mode map at 6 frequency channels of proposed future CMB mission PICO, with fiducial tensor-to-scalar ratios $0.05$ and $0.01$. We present our method's performance to reconstruct the CMB B-mode map, CMB B-mode angular power spectrum, and power spectrum delensing in the following.
\subsection{Cleaned Maps}

\begin{figure}
	\vspace{1cm}
	\includegraphics[scale=.17]{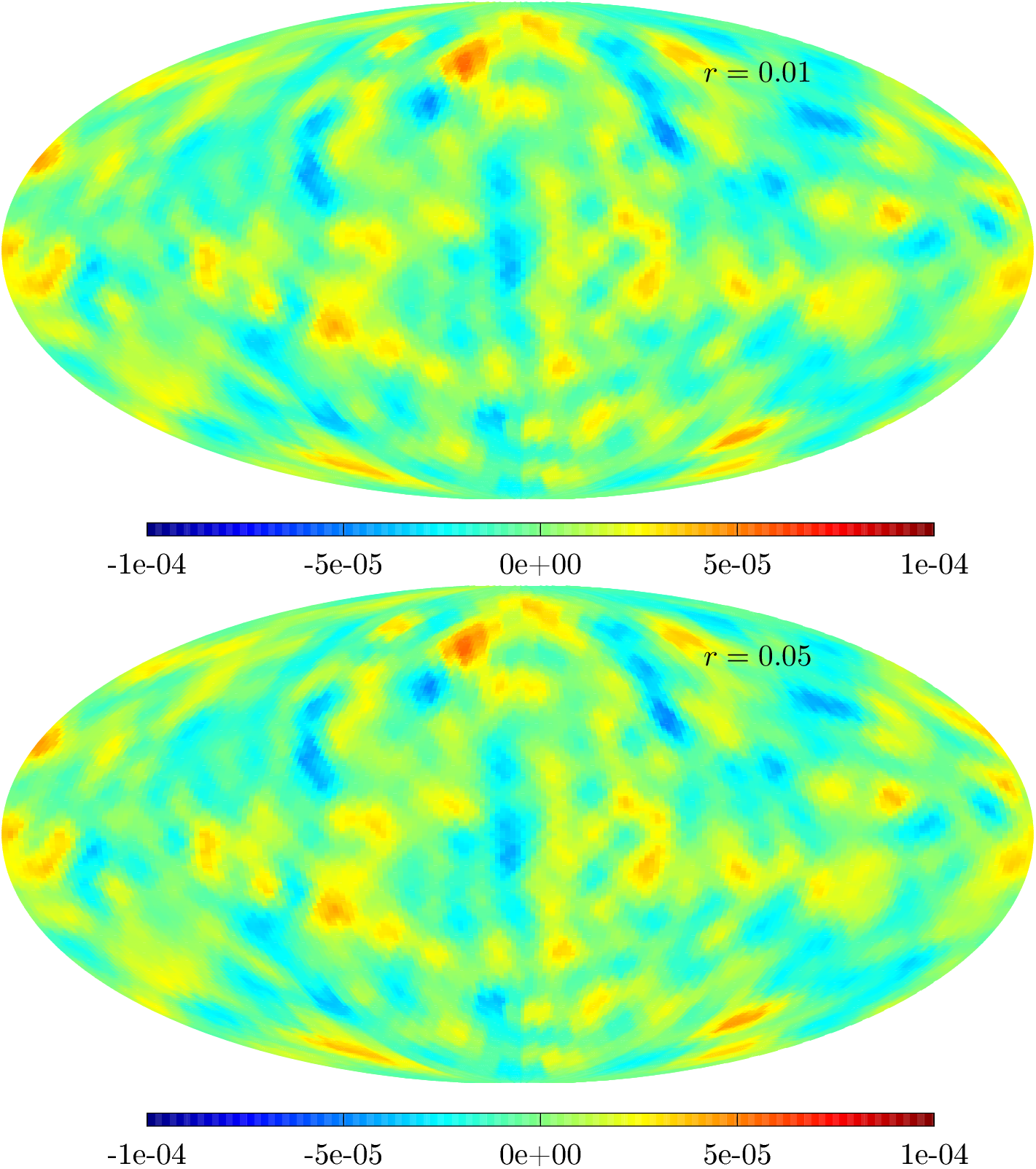}
	\caption{In the above figure, we show the mean of 200 difference maps for $r$ values of 0.01 and 0.05 in the top and bottom panels, respectively. We find a mean map using 99500 map samples from a given simulation, and subtract the input map to obtain the difference map corresponding to the simulation. From the above figure, we find that the mean over simulations of absolute pixel reconstruction error is $\le 10^{-5} \mu K$ for both values of $r$, indicating accurate map reconstruction using our method. The unit is in $\mu K$ thermodynamic.}
	\label{fig:mdiffmap-crop}
\end{figure}

In this subsection, we present the performance of our method to reconstruct the CMB B-mode maps.
In the Figure (\ref{fig:sdmulti_r0p05-crop}) and (\ref{fig:sdmulti_r0p01-crop}) we show pixel standard deviation maps obtained using 200 CMB signal reconstruction following our method for $r=0.05$ and $r=0.01$ respectively. The second last map at the right bottom corner, labeled MEAN, in both the figures shows the mean of all the 200 standard deviation maps obtained from the 200 simulations. The last map at the right bottom corner, labeled STDEV, of both the figures shows the standard deviation maps obtained using the 200 standard deviation maps from the 200 simulations. From the mean, standard deviation map for both the cases, we find reconstruction bias along the galactic plane is $\le 10^{-3}\mu K$. From the standard deviation maps obtained using the 200 standard deviation maps, we find small variation of order $\le 10^{-6} \mu K$ in pixel reconstruction error from one simulation to another for both cases of tensor-to-scalar ratios. In Figure (\ref{fig:mdiffmap-crop}), we show the mean of 200 difference maps for both $r$ values. We find a mean map using 99500 samples of the map from a given simulation, and subtract the input map to obtain the difference map corresponding to the simulation. From the mean difference maps, we find that the mean absolute pixel reconstruction error is $\le 10^{-5} \mu K$ for both values of r, which indicates accurate signal reconstruction using our method.

\subsection{Angular Power Spectrum}
\begin{figure*}
	\hspace{1.5cm}
	\includegraphics[scale=.45]{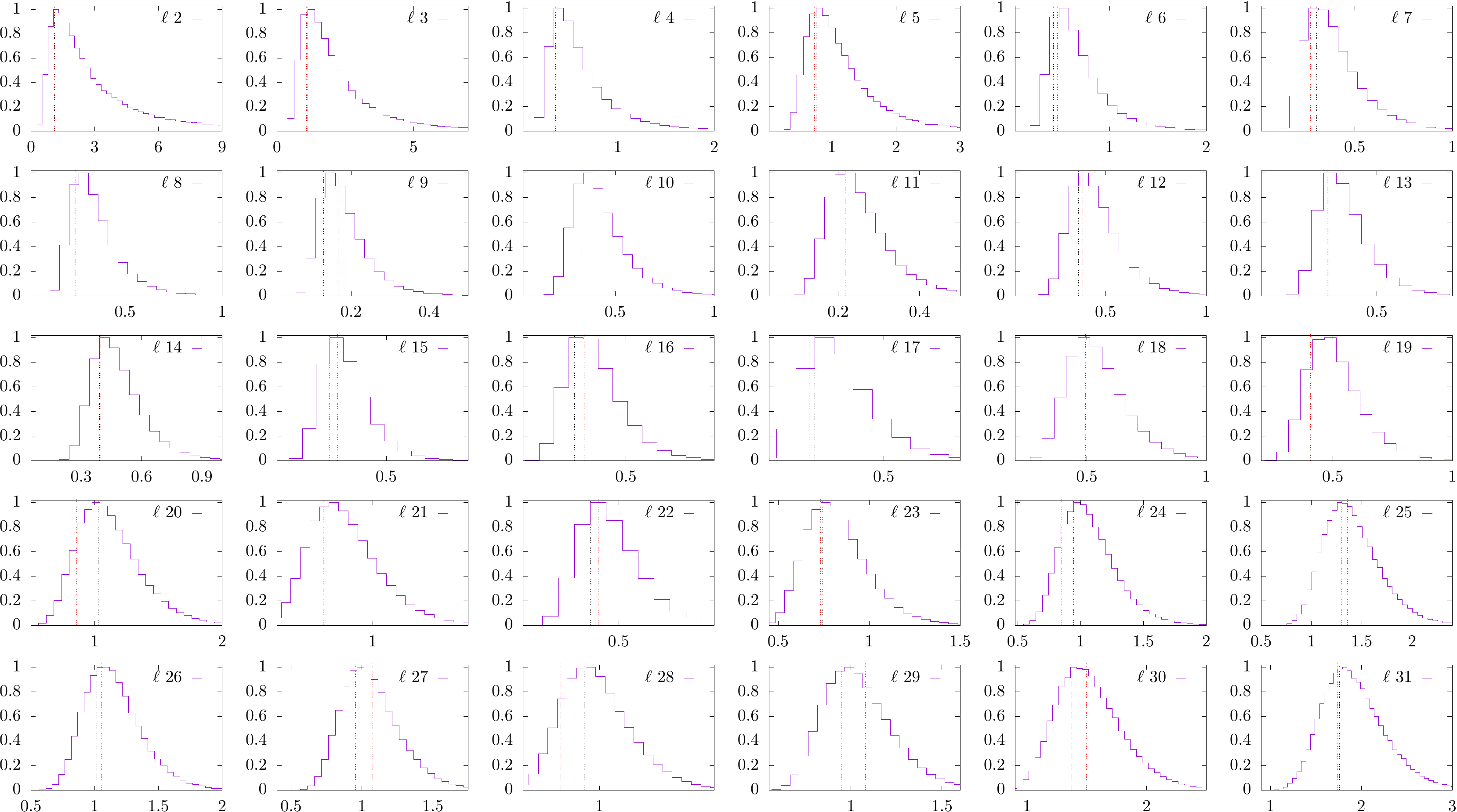}
	\caption{In the above figure, we show normalized densities of the sampled CMB theoretical angular power spectrum obtained by Gibbs sampling for 2 to 31 multipoles for simulation seed 1, with $r=0.05$. The horizontal axis for each subplot represents $\ell(\ell +1)C_{\ell}/2\pi$ in the unit of $10^{-3} \mu K^{2}$. The above histogram gives us the best estimates of the theoretical $C_{\ell}$ (vertical black dashed line) given the data. The vertical red dashed line is the value corresponding to input sky $C_{\ell}$. The position of most of the histogram peeks agree well with the input sky $C_{\ell}$. The deviation of the input $C_{\ell}$ from the peeks in some of the histograms is due to presence of the residual detector noise along with the CMB. The above plots confirms the expected behavior of the $C_{\ell}$ histograms at both low and high $\ell$.  }
	\label{fig:hist_mlt_r0p05-crop}
\end{figure*}

\begin{figure*}
	\hspace{1.7cm}
	\includegraphics[scale=1.3]{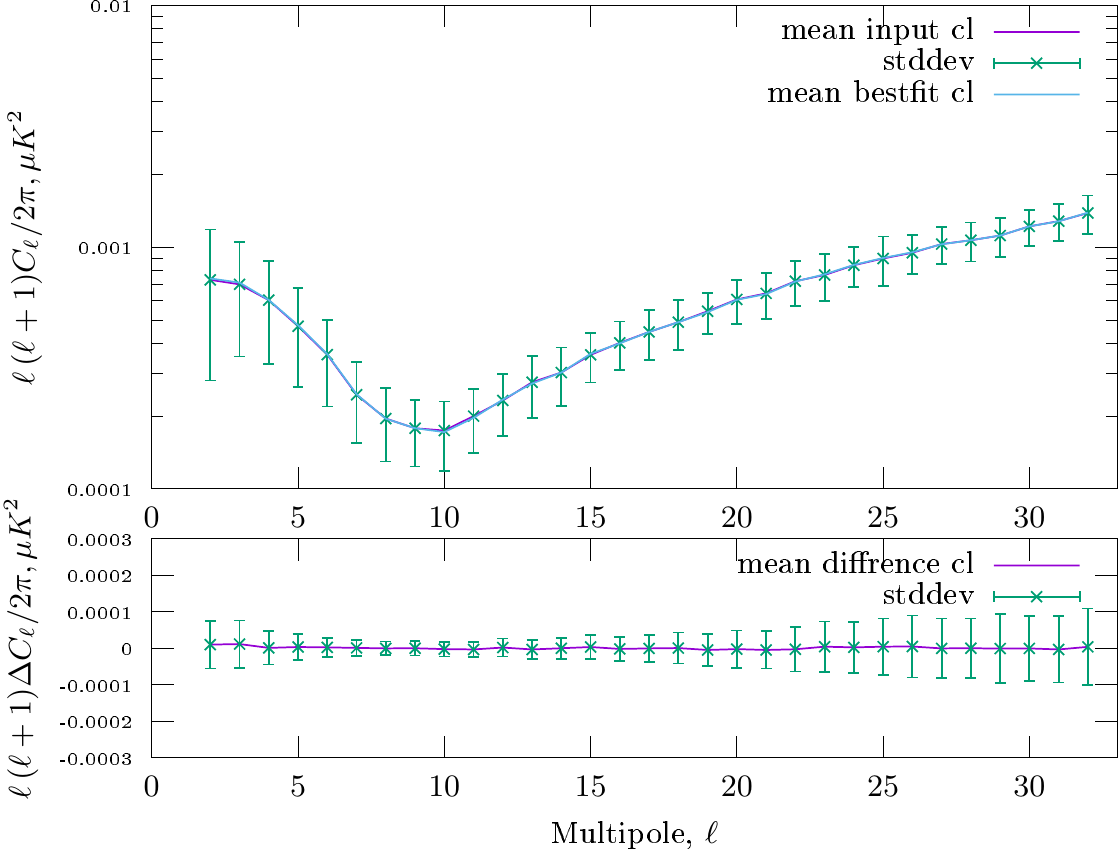}
	\caption{For the tensor-to-scalar ratio $r=0.05$, we plot the mean input angular power spectrum, mean best-fit angular power spectrum, and corresponding standard deviation for 200 simulations in the top panel figure. We show the mean of 200 differences angular power spectrum and associated errors in the bottom panel figure. To get the difference angular power spectrum for a given simulation, we subtract the corresponding input power spectrum from the best-fit angular power spectrum. We find that the mean input and the mean best-fit power spectrum agree very well from the top panel figure. In the bottom panel figure, the mean over simulations of absolute power reconstruction error at each multipole is $< 2\times 10^{-5} \mu K^{2}$. The small absolute reconstruction error at each multipole shows that our method performs very well in reconstructing the angular power spectrum.}
	\label{fig:meancl_sd_r0p05-crop}
\end{figure*}

\begin{figure*}
	\hspace{1.5cm}
	\includegraphics[scale=.45]{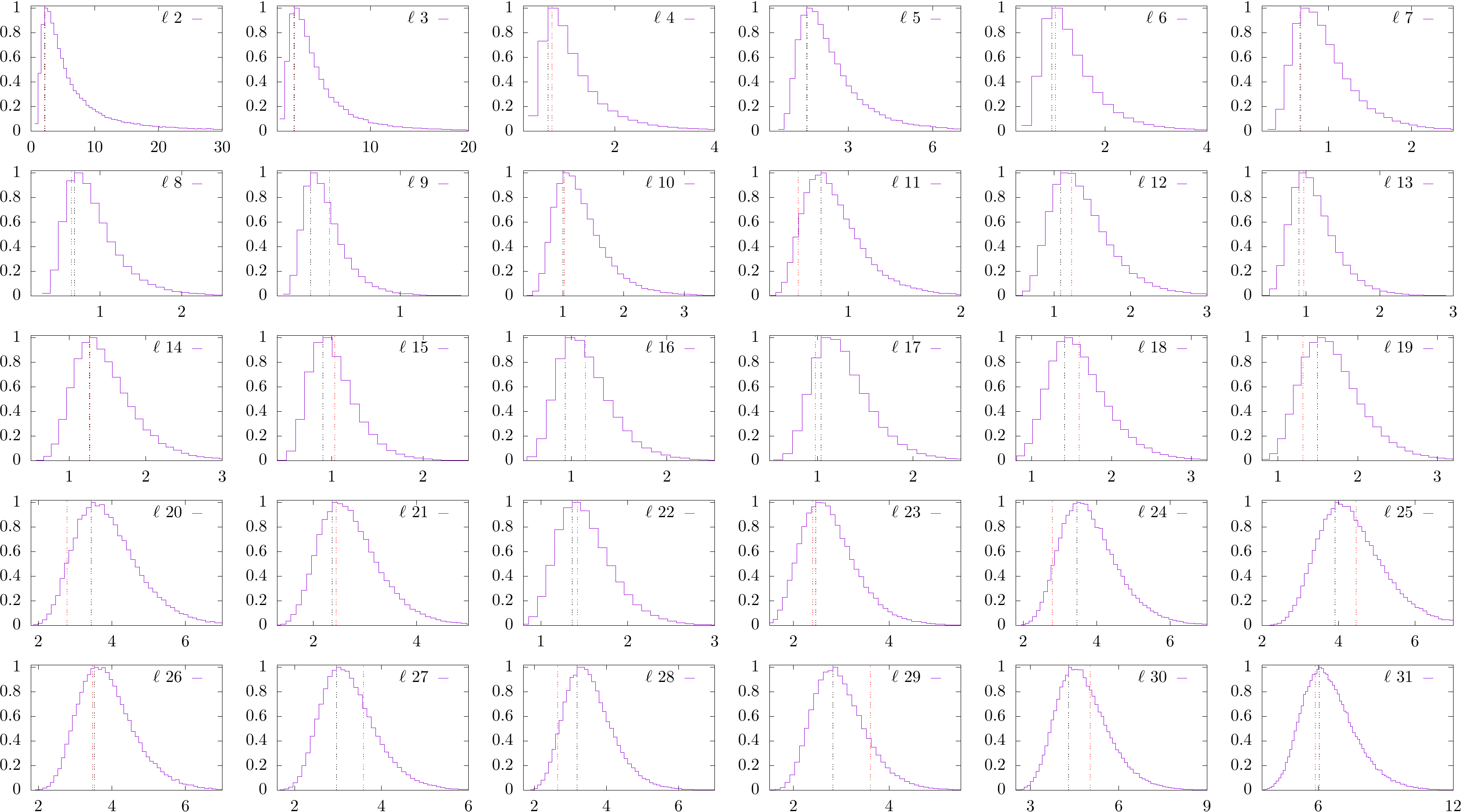}
	\caption{The figure shows normalized densities of the sampled CMB theoretical angular power spectrum obtained by Gibbs sampling for 2 to 31 multipoles for simulation seed 1, with $r=0.01$. The horizontal axis for each subplot represents $\ell(\ell +1)C_{\ell}/2\pi$ in the unit of $10^{-4} \mu K^{2}$. The vertical black dashed line is the best-fit value, and the vertical red dashed line is the value corresponding to input $C_{\ell}$. The position of most of the histogram peeks agree well with the input sky $C_{\ell}$. The deviation of the input $C_{\ell}$ from the peeks in some of the histograms is due to presence of the residual detector noise along with the CMB. The above plots confirms the expected behavior of the $C_{\ell}$ histograms at both low and high $\ell$.}
	\label{fig:hist_mlt_r0p01_crop}
\end{figure*}

\begin{figure*}
	\hspace{1.7cm}
	\includegraphics[scale=1.3]{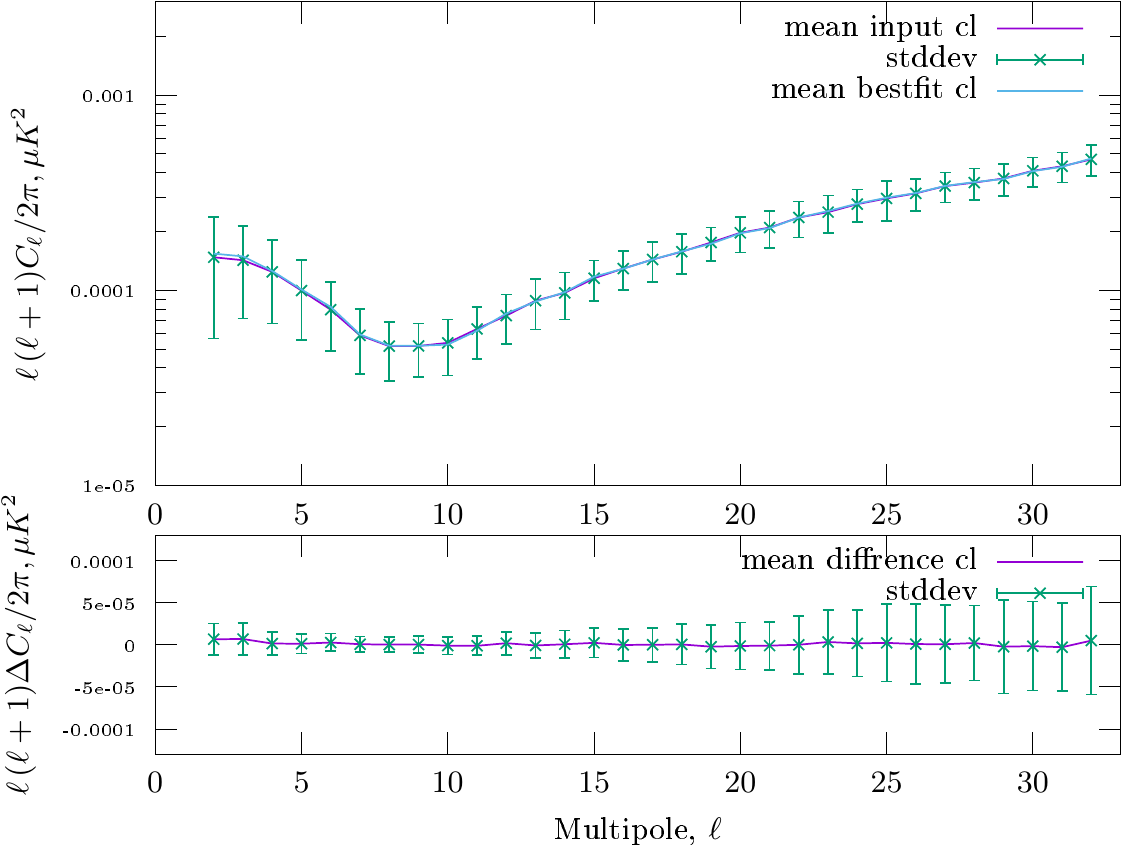}
	\caption{In the top panel figure, we show the mean input angular power spectrum, mean best-fit angular power spectrum, and corresponding error bars. We show the mean of 200 difference angular power spectrum in the bottom panel figure and corresponding error bars. To get the difference angular power spectrum for a given simulation, we subtract the corresponding input power spectrum from the best-fit angular power spectrum. We find the mean best-fit power spectrum from the top panel figure is slightly more than the mean input power spectrum at multipoles $< 7$. In the bottom panel figure, the mean over simulations of absolute power reconstruction error at each multipole is $< 7\times 10^{-6} \mu K^{2}$. The small absolute reconstruction error at each multipole shows that our method performs very well in reconstructing the angular power spectrum, and there is no significant bias in angular power spectrum reconstruction for $r=0.01$. }
	\label{fig:meancl_sd_r0p01-crop}
\end{figure*}

\begin{figure*}
	\hspace{1.3cm}
	\includegraphics[scale=1.3]{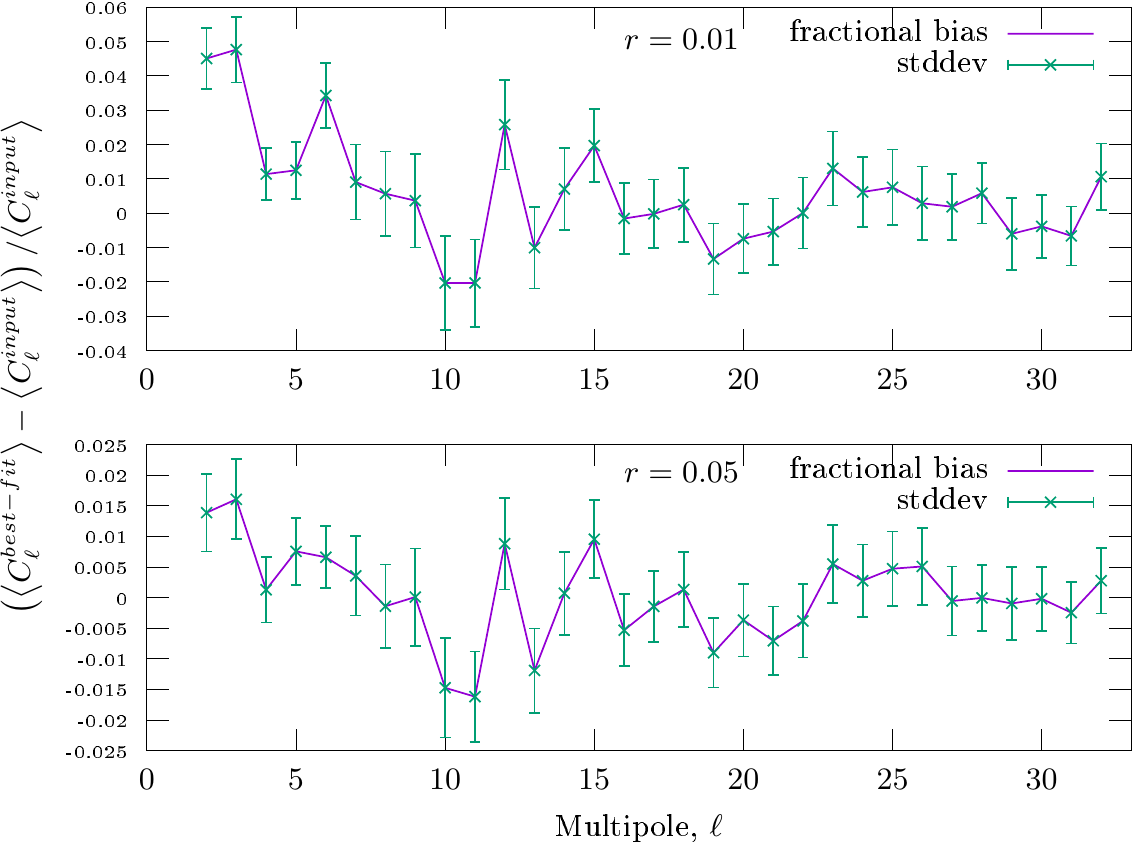}
	\caption{In the top panel of this figure we show the fractional bias corresponding to reconstructed angular power spectrum at each multipole for fiducial tensor-to-scalar ratio of 0.01. In the bottom panel we present the same bias measure for fiducial tensor-to-scalar ratio of 0.05. From the above two plots we find that we have $2\%$ to $3\%$ more positive bias for $r=0.01$ than $r=0.05$, indicating that our method do not have significant bias even for $r=0.01$. This shows that our method performs very well in reconstructing the CMB B-mode angular power spectrum for both the cases. }
	\label{fig:fracbias-crop}
\end{figure*}

In this subsection, we present our method's performance to reconstruct the CMB B-mode angular power spectrum.
We present normalized densities of the Gibb's samples of CMB theoretical angular power spectrum from multipole 2 to 31, along with the input angular power spectrum (vertical black dashed line) and best-fit angular power spectrum (vertical red dashed line) for a randomly chosen simulation seed 1, with $r=0.05$ in Figure (\ref{fig:hist_mlt_r0p05-crop}). In the figure the position of most of the histogram peeks agree well with the input sky angular power spectrum. The deviation of the input angular power spectrum from the peeks in some of the histograms is due to presence of detector noise in contaminated CMB frequency channel maps. The plots in the Figure (\ref{fig:hist_mlt_r0p05-crop}) confirms the expected behavior of the $C_{\ell}$ histograms at both low and high multipoles. For the tensor-to-scalar ratio of 0.05, we present in the top panel of Figure (\ref{fig:meancl_sd_r0p05-crop}) mean over 200 simulations of input angular power spectrum and best-fit angular power spectrum along with corresponding standard deviations to quantify the reconstruction error in CMB B-mode angular ~power spectrum. We also plot in the bottom panel
of the Figure (\ref{fig:meancl_sd_r0p05-crop}), the mean over 200 simulations of difference between best-fit and input angular power spectrum along with corresponding standard deviations to further quantify the reconstruction error in recovered CMB B-mode angular power spectrum. From the upper panel in the Figure (\ref{fig:meancl_sd_r0p05-crop}) we find that the mean input and the mean best-fit power spectrum agree very well for $r=0.05$. From the bottom panel of the Figure (\ref{fig:meancl_sd_r0p05-crop}) we find that the mean over simulations of absolute power reconstruction error at each multipole is $< 2\times 10^{-5} \mu K^{2}$. Similarly we present normalized densities of Gibb's samples of the theoretical angular power spectrum from multipole 2 to 31, along with the input angular power spectrum (vertical black dashed line) and best-fit angular power spectrum (vertical red dashed line), for simulation seed 1, with $r=0.01$ in Figure (\ref{fig:hist_mlt_r0p01_crop}). The best-fit  theoretical angular power spectrum estimate well the input power spectrum. The Figure (\ref{fig:hist_mlt_r0p01_crop}), confirms the expected behaviour of the angular power spectrum histograms at both low and high multipoles. In the Figure (\ref{fig:meancl_sd_r0p01-crop}) for $r=0.01$, we present in the top panel, mean over 200 simulations of input and best-fit angular power spectrum, in the bottom panel, the mean over 200 simulations of difference between the best-fit angular power spectrum and input angular power spectrum along with corresponding standard deviations. From the upper panel in Figure (\ref{fig:meancl_sd_r0p01-crop}), we find that the mean best-fit power spectrum has more power than the mean input power spectrum at multipoles $< 7$. From the bottom panel of the Figure (\ref{fig:meancl_sd_r0p01-crop}) we find that the mean over simulations of absolute power reconstruction error at each multipole is $< 7\times 10^{-6} \mu K^{2}$ for $r=0.01$. We plot in the Figure (\ref{fig:fracbias-crop}) the fractional bias $\Delta C_{\ell}^{fb}$ in recovered angular power spectrum calculated using 200 best-fit angular power spectrum and input angular power spectrum, defined as:
\begin{equation*}
\Delta C_{\ell}^{fb} = \frac{\left\langle C_{\ell}^{best-fit} \right\rangle - \left\langle C_{\ell}^{input} \right\rangle}{\left\langle C_{\ell}^{input} \right\rangle}
\label{eq:fbais}
\end{equation*}

From the plot in the Figure (\ref{fig:fracbias-crop}), we find $2\%$ to $3\%$ more bias in the reconstructed power spectrum for simulated CMB B-mode maps with tensor-to-scalar ratio $0.01$ than $0.05$, indicating that our method does not have significant bias even when $r=0.01$. This shows that our method performs very well in reconstructing the CMB B-mode angular power spectrum for both the cases.  


\subsection{Reconstructing $r$}
Using set of Gibbs samples $C_{\ell}^{B~i}$  and Blackwell-Rao Estimator  \citep{2005PhRvD..71j3002C} we in a self-consistent Bayesian framework compute the joint posterior distribution $P(r,A^{\textrm{lens}})$ of the tensor-to-scalar ratio $r$ and the amplitude of lensing, $A^{\textrm{lens}}$. To estimate the cosmological parameter $r$ and $A^{\textrm{lens}}$ we maximize the likelihood 
\begin{equation}
\mathcal{L}(C_{\ell}^{B}|C_{\ell}^{B~th}) = \exp\left(-\frac{1}{2}\sum_{\ell=2}^{{\ell}max}(2\ell+1)\left[\ln\left(\frac{C_{\ell}^{B~th}}{C_{\ell}^{B}}\right)+\frac{C_{\ell}^{B}}{C_{\ell}^{B~th}}-1\right] \right)
\label{eq16}
\end{equation}
where the model theoretical CMB B-mode power spectrum $C_{\ell}^{B~th}$ is given by linear sum 
\begin{equation}
C_{\ell}^{B~th} = \left(\frac{r}{0.05}\right)C_{\ell}^{T}(r=0.05) + A^{\textrm{lens}}C_{\ell}^{L}(r=0)
\label{eq17}
\end{equation} 
where $C_{\ell}^{T}(r=0.05)$ is the tensor B-mode power spectrum, for a tensor-to-scalar ratio $r=0.05$, and $C_{\ell}^{L}(r=0)$ is the lensing-induced B-mode power spectrum. In order to estimate the joint posterior distribution of r and $A^{\textrm{lens}}$, $P(r,A^{\textrm{lens}})$ we vary both r and $A^{\textrm{lens}}$ and make use of the Blackwell-Rao approximation
\begin{equation}
P(r,A^{\textrm{lens}}) \approx \frac{1}{N}\sum_{i=1}^{N}\mathcal{L}(C_{\ell}^{B~i}|C_{\ell}^{B~th}(r,A^{\textrm{lens}}))P^{prior}(A^{\textrm{lens}})
\label{eq18}
\end{equation}
where N is the total number of Gibbs samples of $C_{\ell}^{B~i}$ used. For large N the Blackwell-Rao estimate becomes an exact approximation of $P(r,A^{\textrm{lens}})$ \citep{2005PhRvD..71j3002C}. In this work, we do not put any prior on $A^{\textrm{lens}}$ so that $P^{prior}(A^{\textrm{lens}})$ is a constant.
\hspace{.7cm}

\begin{figure*}
	\hspace{.5cm}
	\includegraphics[scale=0.6]{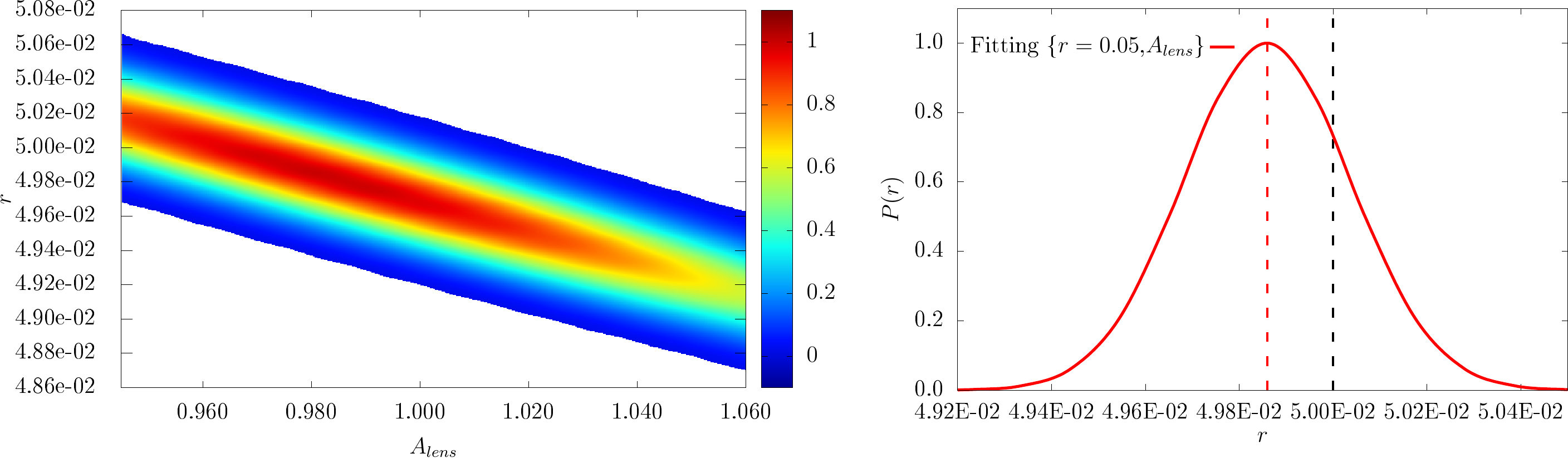}
	\caption{In the left panel figure, we plot normalized joint 2-D Blackwell-Rao posterior density estimates for $A^{\textrm{lens}}$ and $r$  from 200 different simulations for the case with fiducial tensor-to-scalar ratio 0.05. The color box shows the range of normalized likelihood. In the right panel figure, we show the posterior distribution of $r$, which we get by slicing the joint 2-D Blackwell-Rao posterior density on the left, for maximum likelihood of $A^{\textrm{lens}}$. The vertical black dashed line represents the fiducial tensor-to-scalar ratio of the input CMB B-mode maps, and the vertical red dashed line represents the mode of the r posterior. We can see in the right panel figure that the fiducial $r$ value is within the $1\sigma$ of the posterior, which indicates that we can achieve significant delensing of $P(r)$ using our method.}
	\label{fig:L_vs_rA_0p05-crop}
\end{figure*}

In the following, we discuss our Bayesian method's performance to delens the CMB B-mode angular power spectrum and hence minimize the lensing contribution to the recovered distribution of the tensor-to-scalar ratio $P(r)$. 
For fiducial tensor-to-scalar ratio 0.05 we plot in the Figure (\ref{fig:L_vs_rA_0p05-crop}) normalized joint 2-D Blackwell-Rao posterior density estimates $P(r,A^{\textrm{lens}})$ along with the normalized posterior distribution $P(r)$. We get the $P(r)$ by slicing the joint 2-D Blackwell-Rao posterior density $P(r,A^{\textrm{lens}})$, for maximum likelihood of $A^{\textrm{lens}}$, using set of Gibb's samples $\{C_{\ell}^{B,i}\}$ for each of the 200 different simulations. Since the true value 0.05 is within $1\sigma$ of the normalized posterior, we conclude that our method performs well in reconstructing the angular power spectrum for $r=0.05$ hence delensing the angular power spectrum. Similarly we plot normalized joint 2-D Blackwell-Rao posterior density $P(r,A^{\textrm{lens}})$ in left panel and  the posterior distribution $P(r)$ in the right panel of the Figure (\ref{fig:L_vs_rA_0p05-crop}) for fiducial tensor-to-scalar ratio 0.01. We find that the true value 0.01 is within $1\sigma$ of the normalized posterior $P(r)$, establishing that our method also performs well in reconstructing the angular power spectrum for $r=0.01$ and delensing the angular power spectrum.

\begin{figure}
	\hspace{.0cm}
	\includegraphics[scale=.9]{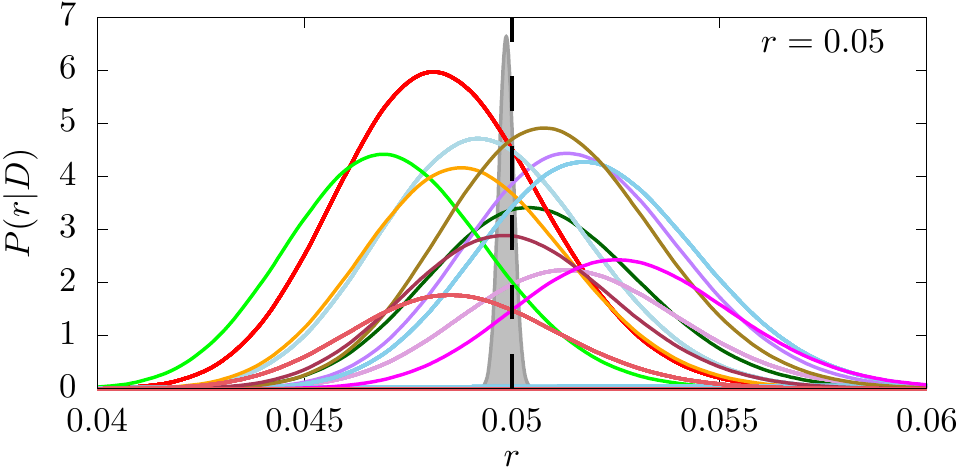}
	\caption{Blackwell-Rao posteriors from each of randomly selected 12 different simulations for tensor-to-scalar ratio 0.05. The gray band is the product of the posteriors from all the 200 simulations with the tensor-to-scalar ratio 0.05. We arbitrarily normalize the posteriors to fit on these axes. The dark black dashed vertical line represents the simulated fiducial tensor-to-scalar ratio. We expect that the gray band covers the fiducial value of tensor-to-scalar ratio 0.05 to within its width, as is indeed the case showing that the posterior $P(r|D)$ converges.}
	\label{fig:prob_various_0p05-crop}
\end{figure}
\begin{figure}
	\hspace{-0.cm}
	\includegraphics[scale=.9]{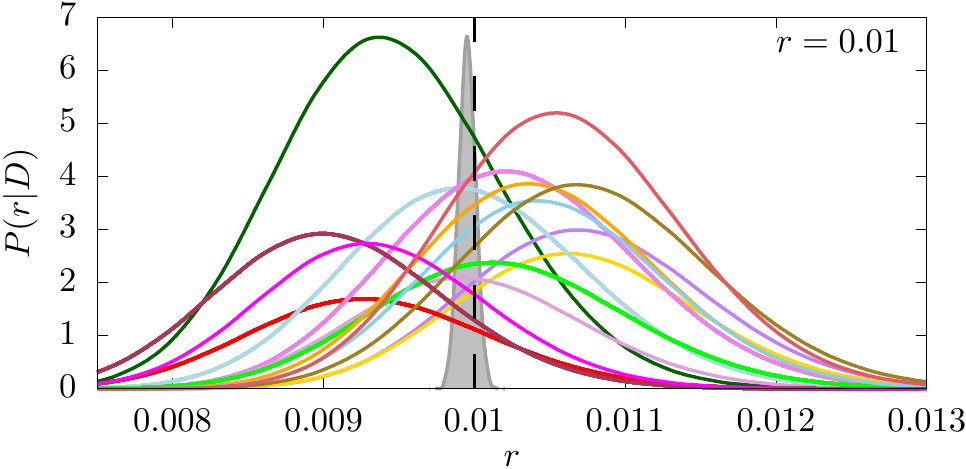}
	\caption{In the above plot, we show Blackwell-Rao posteriors from randomly selected 15 simulations for the fiducial tensor-to-scalar ratio 0.01. The gray band is the product of the posteriors from all the 200 simulations for the tensor-to-scalar ratio 0.01. We arbitrarily normalize the posteriors to fit on these axes. The dark black dashed vertical line represents the simulated fiducial tensor-to-scalar ratio. We expect that the gray band covers the fiducial value of the tensor-to-scalar ratio 0.01 to within its width, as is indeed the case. This shows that the posterior $P(r|D)$ converges for $r=0.01$.}
	\label{fig:prob_various_0p01-crop}
\end{figure}

To show convergence of the posterior $P(r|D)$  we show the product of the 200 posteriors in Figure (\ref{fig:prob_various_0p05-crop}) and Figure (\ref{fig:prob_various_0p01-crop}) as shaded gray band for fiducial tensor-to-scalar ratio 0.05 and 0.01 respectively. Since the fiducial value of $r$ in both the cases is within the corresponding gray band's width, we conclude that chains converge, and our method is correct. We also show posterior for some randomly chosen simulations, which we normalize arbitrarily in each of the plots to fit on axes.  

\begin{figure*}
	\hspace{0.2cm}
	\includegraphics[scale=.6]{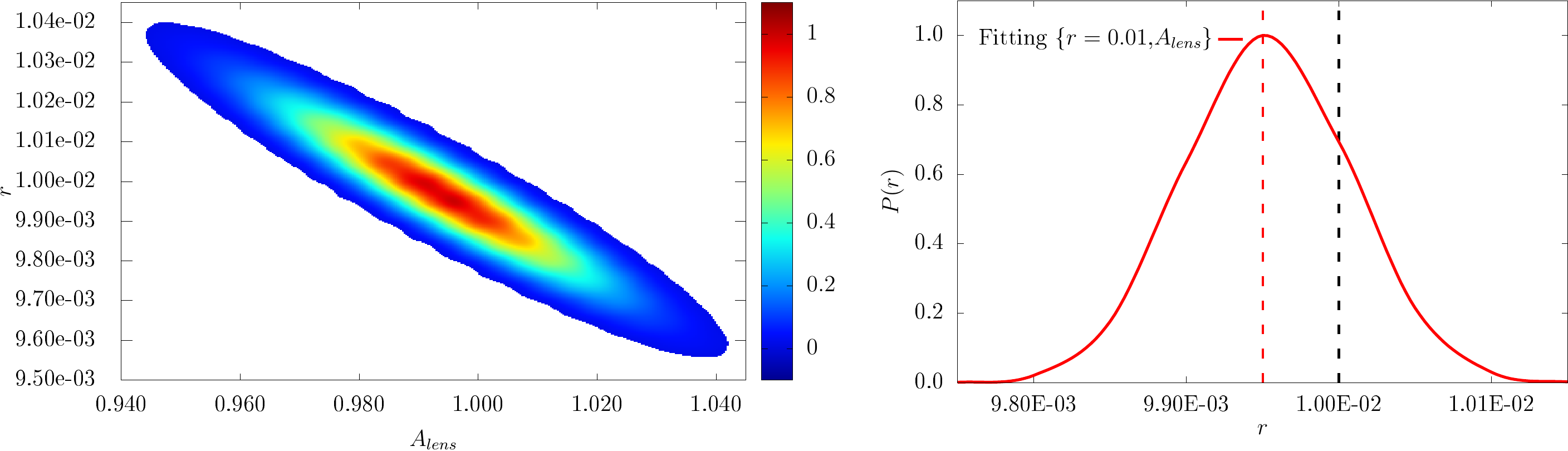}
	\caption{The left panel figure shows normalized joint 2-D Blackwell-Rao posterior density estimates for $A^{\textrm{lens}}$ and $r$ from 200 different simulations for the case with fiducial tensor-to-scalar ratio 0.01. The color box shows the range of normalized likelihood. In the right panel figure, we show the posterior distribution of $r$, which we get by slicing the joint 2-D Blackwell-Rao posterior density on the left, for maximum likelihood of $A^{\textrm{lens}}$. The vertical dashed line represents the fiducial tensor-to-scalar ratio of the input CMB B-mode maps, and the vertical red dashed line represents the r corresponding to the maximum likelihood for the posterior distribution.  We can see in the right panel figure that the fiducial $r$ value is within the $1\sigma$ of the posterior, which indicates that we achieve significant delensing of $P(r)$  using our method.}
	\label{fig:L_vs_rA_0p01-crop}
\end{figure*}
	\vspace{1.5cm}

\vspace{-1cm}
\section{Discussions \& Conclusion}
\label{Diss}
We develop a new foreground model-independent approach to measure CMB B-mode signal and angular power spectrum using simulated observations for proposed future generation, PICO satellite mission in this work.  Our new non-parametric method is useful since
spatial and spectral variations of the polarized foreground component may not be accurately known. In this article, we extend and improve the earlier reported Bayesian ILC method, to reconstruct weak CMB B-mode signals by introducing noise bias corrections at two stages during the ILC weight estimation. We extensively test our new method's performance to reconstruct the CMB B-mode sky signal and angular power spectrum and obtain the joint distribution of tensor to scalar ratio and lensing amplitude for two different values of $r$. The proposed future generation CMB B-mode mission, like PICO, can break the power spectrum degeneracy between primordial B-mode and lensing B-mode by detecting the reionization bump. Utilizing this advantage, we further perform the delensing of the recovered CMB B-mode angular power spectrum. We use Gibb's samples of the CMB B-mode theoretical angular power spectrum obtained from our method to separate the primordial and lensing B-mode contribution to the recovered distribution of the tensor-to-scalar ratio in a Bayesian manner and have quantified the performance of our method at two different tensor-to-scalar ratios.

We summaries the findings of our method in the following :
\begin{enumerate}
		
	\item From the mean, standard deviation maps for $r=0.05$ and $0.01$, we find reconstruction bias along the galactic plane is $\le 10^{-3}\mu K$. From the standard deviation over simulations of standard deviation, we find small variation of order $\le 10^{-6} \mu K$ in pixel reconstruction error from one simulation to other for both  $0.05$ and $0.01$ tensor-to-scalar ratios. We also find the mean over simulations of absolute pixel reconstruction error is very small ($\le 10^{-5} \mu K$) for both the cases. In light of the above, we conclude that our method accurately reconstructs the simulated primordial CMB B-mode sky for both $r=0.05$ and $r=0.01$ cases.  
	
	\item We find the mean input power spectrum and the mean best-fit power spectrum agree very well for $r=0.05$ case, whereas for $r=0.01$, there is slightly more mean power in the reconstructed power spectrum at multipoles $ < 7$. Using fractional bias to quantify this positive bias, we find it to be only $2\%$ to $3\%$ more for the case with $r=0.01$ than $r=0.05$, which indicates that our method does not have significant bias even for   $r=0.01$ and performs very well in reconstructing the angular power spectrum for both the cases.
	
	\item In this work, we estimate the joint posterior for the CMB B-mode signal and its theoretical angular power spectrum over the large angular scales. We also obtain the appropriate confidence intervals for the theoretical angular power spectrum necessary for cosmological parameter estimation. This is the first demonstration of reconstruction of the CMB B-mode signal using an ILC approach following a Bayesian framework.
	
	

	\item On fitting both $r$ and $A^{\textrm{lens}}$, we find the fiducial tensor-to-scalar values to be within $1\sigma$ of the recovered distribution $P(r)$ for both the cases. This shows that the samples of the Gibb's CMB  B-mode theoretical angular power spectrum, obtained using our method, gives an unbiased estimate of $P(r)$ for both the cases. The power spectrum delensing method used in this article cannot remove the lensing B-mode cosmic variance induced by E-modes. However, using our new method on the foreground and noise-contaminated six PICO CMB B-mode channels, we can detect $r$ with more than $9\sigma$ and $8\sigma$ significance if the true values of $r$ were $0.05$, $0.01$ respectively at low resolutions without using any $A^{\textrm{lens}}$ prior. 
	
	\item Our method does not explicitly require foreground model. Thus any error due to incorrect foreground B-mode model, does not bias our results.
	
	\item Our method is computationally fast, efficient, and accurate in delensing and detecting significant unbiased detection for levels of $r\ge 10^{-2}$. 
	
\end{enumerate}

\section*{Acknowledgments}

 We use the publicly available HEALPix~\citep{2005ApJ...622..759G} package available to perform spherical harmonic decomposition and for visualization purposes. 
  We acknowledge the use of the Legacy Archive for Microwave Background Data Analysis (LAMBDA) and Planck Legacy Archive (PLA). LAMBDA is a part of the High Energy Astrophysics Science Archive Center (HEASARC) and the Planck Legacy Archive (PLA) contains all public products originating from the Planck mission, an ESA science mission with instruments and contributions directly funded by ESA Member States, NASA, and Canada. This research has made use of NASA's Astrophysics Data System.

\end{document}